\documentclass[aps,prd,twocolumn,showpacs,preprintnumbers,amsmath,amssymb]{revtex4-1}

 \def\ep{{\epsilon}}

 \def\frac#1#2{{#1\over #2}}

 \def\s{\sqrt}

\def\be{\begin{equation}}
\def\ee{\end{equation}}
\def\ba{\begin{eqnarray}}
\def\ea{\end{eqnarray}}

 \def\de{\partial}

 \def\f {\frac}
 \def\ti{\tilde}
 \def\ap{\alpha}

 \def\la{\langle}
 \def\lb{\rangle}
 \def\ep{\epsilon}

\usepackage{color}
\usepackage{graphicx}
\usepackage{dcolumn}
\usepackage{bm}
\usepackage[hypertex]{hyperref}
\usepackage{CJK}

\begin{document}
\begin{CJK*}{UTF8}{bsmi}

\title{Holography and Entanglement in Flat Spacetime}

\author{Wei Li (\CJKfamily{gbsn}李~微) 
and Tadashi Takayanagi (\CJKfamily{min}高柳~匡)
}
\affiliation{
Institute for the Physics and Mathematics of the Universe (IPMU),
University of Tokyo, Kashiwa, Chiba 277-8582, Japan
            }

\date{\today}

\begin{abstract}
We propose a holographic correspondence of the flat spacetime based on the behavior of the entanglement entropy and the correlation functions. The holographic dual theory turns out to be highly nonlocal. We argue that after most part of the space is traced out, the reduced density matrix gives the maximal entropy and the correlation functions become trivial. We present a toy model for this holographic dual using a nonlocal scalar field theory that reproduces the same property of the entanglement entropy. Our conjecture is consistent with the entropy of Schwarzschild black holes in asymptotically flat spacetimes.
\end{abstract}

\maketitle
\end{CJK*}

\noindent
{\bf Introduction}.
One of the most powerful tools to study quantum gravity is the holographic duality conjecture: the quantum gravity in spacetime $\cal{M}$ is equivalent to a quantum field theory living on the boundary $\partial\cal{M}$ \cite{tH}. In particular, the quantum
gravity in anti-de Sitter space has been well developed via the
AdS/CFT correspondence \cite{Maldacena} from string theory. However, to
understand our universe, we
need to study quantum gravity in other spacetimes such as flat
space, de Sitter space, and the big bang spacetime. The purpose of
this Letter is to investigate holography in flat spacetime and to
present a consistent outline of its basic properties and mechanism.
We will focus on the Euclidean flat spacetime $\mathbb{R}^{d+1}$ since the Euclidean formulation is often simpler and better defined than the Lorentzian version, as in AdS/CFT
\cite{GKP}.

In polar coordinates, the metric of the Euclidean spacetime $\mathbb{R}^{d+1}$ is
\begin{equation}\label{metric}
ds^2_{\mathbb{R}^{d+1}}=d\rho^2+\rho^2 ds^2_{S^d}
\end{equation}
The holographic principle dictates that the boundary dual theory of the gravity in $\mathbb{R}^{d+1}$ lives on the unit-radius sphere $S^d$ at $\rho=\rho_\infty$, where
$\rho_{\infty}$ is the bulk cutoff radius and is related to the UV cutoff in boundary field theory (as in AdS/CFT); we take the limit $\rho_\infty\to \infty$.

The assumptions we adopt in this Letter are the following: (1) the dual field theory
allows a path-integral formulation even if it is nonlocal; (2) the bulk-to-boundary correspondence holds, i.e., the partition function of gravity in $\mathbb{R}^{d+1}$ equals that of holographic dual theory on $S^{d}$ \cite{GKP}. In Lorentzian holography, $S^d$ is replaced by the
$d$-dimensional de Sitter space (see \cite{Marolf,dSdS} for
earlier studies). See
\cite{BS,Sus,St} for other approaches to holography in flat
space.

\noindent
{\bf Holographic entanglement entropy.}
When a quantum system is divided into two subsystems $A$ and $B$, the von Neumann entropy $S_A=-\mbox{Tr}[\rho_A\mbox{log}\rho_A]$ (where $\rho_A$ is the reduced density matrix after tracing out $B$) is called the entanglement entropy of the subsystem $A$.
The scaling behavior and certain universal coefficients of the entanglement entropy encode important information on the degrees of freedom and nonlocal correlations of the system.
Furthermore, the entanglement entropy is a general-purpose quantity since it can be defined in any quantum many-body system that allows a path-integral formalism --- even in nonlocal field theories, as will be shown later. Thus the entanglement entropy is particularly useful when we know little else about the holographic dual of a given gravity theory as in our case.

On the gravity side, there is a general prescription to compute the entanglement entropy holographically: when the $d$-dimensional boundary system is divided into two parts $A$ and $B$, the holographic entanglement entropy of the subsystem $A$ is given by the formula \cite{RT}
\begin{equation}
\label{holoEE}
S_A^{hol.}=\f{\mbox{area}(\gamma_A)}{4G^{(d+1)}_N},
\end{equation}
where $\mbox{area}(\gamma_A)$ is the area of the minimal surface $\gamma_A$ that lies inside the $(d+1)$-dimensional bulk and borders on the boundary $\partial A$ of the subsystem $A$; $G^{(d+1)}_{N}$ is the $(d+1)$-dimensional Newton's constant.

\begin{figure}
 \begin{center}
 \includegraphics[height=2.5cm,clip]{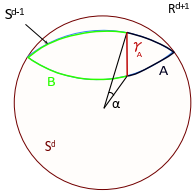}
 \end{center}
 \caption{
 \label{fig}
The geometric computation of the entanglement entropy $S_A$ in the
flat space holography.}
 \end{figure}

Now we apply (\ref{holoEE}) to compute the holographic entanglement entropy of a Euclidean field theory living on the boundary of $\mathbb{R}^{d+1}$.
The metric of the boundary sphere $S^d$ is $ds^2_{S^d}=d\tau^2+\cos^2\tau d\Omega^2_{d-1}$, where $\tau$ $\in [-\frac{\pi}{2},\frac{\pi}{2}]$ is regarded as the Euclidean time, and the spatial slice of constant $\tau$ is $S^{d-1}$, whose metric can be written as
$d\Omega^2_{d-1}=d\theta^2+\sin^2\theta d\Omega^2_{d-2}$. We divide the spatial slice $S^{d-1}$ at $\tau=\tau_0$ into two spherical caps $A$ and $B$ using a subsphere $S^{d-2}$ given by $\theta=\theta_0$. The radius of this $S^{d-2}$ in
$\mathbb{R}^{d+1}$ is $\rho_\infty \cos\tau_0\sin\theta_0\equiv \rho_{\infty} \sin{\frac{\alpha}{2}}$, where $\alpha \in [0,2\pi]$ and $\rho_{\infty}\alpha$ is the geodesic distance in $S^{d}$ of antipodal points of $S^{d-2}$ (see Fig.\ref{fig}).

The holographic
entanglement entropy is
\be
S_A^{hol.}=S_B^{hol.}=\frac{\pi^{\f{d-1}{2}}}{\Gamma(\frac{d+1}{2})}\cdot
\f{\rho^{d-1}_\infty\left(\sin\f{\alpha}{2}\right)^{d-1}}{4G_N^{(d+1)}}.
\label{entt} \ee

First we notice that for a small subsystem A ($\alpha\ll 1$),
(\ref{entt}) approaches $A$'s volume  instead of its area.
Moreover, $S_A$ with generic $\alpha$ is extensive since it is
proportional to the spatial volume of the full boundary system (see
also \cite{Barbon}). These two facts are in
sharp contrast with the behavior both in local field theories \cite{Areal} and
in AdS/CFT. In a local field theory at its ground state, the leading
divergence of the entanglement entropy is always proportional to
the surface area of the subsystem (so-called ``area law") \cite{Areal}. The entropy
becomes extensive only when the system is in highly excited states
with energy around the UV cutoff \cite{CC,AFC}.
However, the holographic dual of the gravity in flat space should not be restricted to any particular type of states in the boundary theory since this ``volume law" applies to the holographic entanglement entropy of any asymptotically flat space. Taking all of these into account, we conjecture that the holographic dual is described by a certain
nonlocal field theory. Below we will construct one such example
based on the scalar field theory.

Also note that in the $\alpha\to 0$ limit $S_A$
saturates the holographic bound \cite{HB} (here given by the volume of $A$ in $S^d$). Therefore although our total system is in a pure state as evidenced by $S_A=S_B$, an infinitesimal subsystem $A$ has a density matrix $\rho_A$ with maximal entropy (=$\log{dim\cal{H}_A}$ where $\cal{H}_A$ is the Hilbert space of $A$).
Namely, an infinitesimal subsystem $A$ is maximally
entangled with its complement $B$.

Let us consider a generic (not necessarily local) free scalar field theory on
$S^d$ defined by the action
\begin{equation}
S_{boundary}=\int d\Omega_d \left[\phi \cdot f(-\Delta)\cdot
\phi\right], \label{actions}
\end{equation}
where $\Delta$ is the Laplacian on $S^d$ and $f(x)$ is a
smooth function (see \cite{Nesterov:2010yi} for an analogous
computation in $\mathbb{R}^d$).

To see the extensive behavior of the entanglement entropy, it suffices to consider the simplest configuration with $\alpha=\pi$. In this case, $S_A$ can be expressed as follows (similar to the geometric entropy in \cite{FNT}):
\begin{equation}
S_A=\f{\de}{\de N}\log
\f{Z_N}{(Z_1)^{1/N}}\Biggr|_{N=1}, \ee where $Z_N$ is the partition
function of (\ref{actions}) on the orbifold $S^d/\mathbb{Z}_N$;
the $\mathbb{Z}_N$ action is defined by a
$\f{2\pi}{N}$ rotation of $S^d$.
The partition function can be evaluated via Schwinger representation: $\log Z_N=\f{1}{2}\int^\infty_{\ep} \f{ds}{s}\mbox{Tr}_{(N)}
e^{-sf(-\Delta)}$, where $\ep$ is related to the UV cutoff in the field theory.

Spherical harmonics on $S^d$ are labeled by angular momenta
$(l,m_1,\dots,m_{d-1})$, which range as $l\geq m_1 \geq \ldots \geq m_{d-2}\geq 0$ and $m_{d-2}\geq |m_{d-1}|$.
The eigenvalues of the Laplacian $\Delta$
are $-l(l+d-1)$. The $\mathbb{Z}_N$ orbifolding acts by multiplying a phase factor $e^{\f{2\pi i}{N}m_{d-1}}$. The relevant trace of kernel $Tr\equiv \mbox{Tr}_{(N)} e^{-sf(-\Delta)}-\f{1}{N}\mbox{Tr}_{(1)}
e^{-sf(-\Delta)}$ is then:
\begin{eqnarray}
Tr=\f{1}{N}\sum_{k=1}^{N-1}\sum_{l}e^{-sf(l(l+d-1))}g\left(l,d,\frac{k}{N}\right),
\end{eqnarray}
where $g(l,d,\frac{k}{N})\equiv\sum_{\{m_i\}}e^{2\pi i\f{k}{N}m_{d-1}}$
incorporates the sum over all magnetic angular momenta $m_i$ and is computed to be
\be
g(l,d,\frac{k}{N})
=\sum^{l}_{n=0}\binom{n+d-3}{d-3}\sum^{l-n}_{m=-(l-n)}e^{2\pi i\f{k}{N}m}
\ee
for $d=2$, the binomial is $\delta_{n,0}$.

Lower dimensional spheres have more compact results: $g(l,2,\frac{k}{N})=\frac{\sin{[\frac{\pi
k}{N}(2l+1)]}}{\sin{(\frac{\pi k}{N})}}$ and $g(l,3,\frac{k}{N})=
\left(\frac{\sin{[\frac{\pi
k}{N}(l+1)]}}{\sin{(\frac{\pi k}{N})}}\right)^2
$ and need to be treated separately.

For higher dimensional spheres ($d\ge4$), $g(l,d,\frac{k}{N})$ is a degree ($d-3$) (pseudo-)polynomial of $l$ with leading term $
\f{l^{d-3}}{2\cdot(d-3)!\cdot\sin^2\f{\pi k}{N}}
$. Summing over all twisted sectors using $\sum_{k=1}^{N-1}\sin^{-2}\f{\pi k}{N}=\f{1}{3}(N^2-1)$ and then applying $\lim_{N\to 1} \frac{\partial}{\partial N}$, we obtain the leading divergence of the entanglement entropy:
\begin{equation}
S^{d\geq4}_A=\f{1}{6}\int^\infty_{\ep}\f{ds}{s}\sum_{l=0}^\infty
\f{l^{d-3}}{(d-3)!}e^{-sf(l(l+d-1))} + \dots.
\end{equation}

Now we impose the UV cutoff. For $S^d$ with radius $L$ and lattice
spacing $a$, the azimuthal angular momentum $l$ has an upper bound
given by $l_{max}=\frac{L}{a}$.
This translates into a lower bound on the integration parameter $s$: $s \geq \epsilon=\frac{1}{f(l^2_{max})}$.

First, let us look at actions with $f(x)=x^p$; in particular, $p=1$ gives the standard massless scalar.
The leading divergence of the entanglement entropy is
\be \label{localleading}S_A =\f{\Gamma(\f{d-2}{2p})}{6\cdot
(d-2)!}~\left(\frac{L}{a}\right)^{d-2}. \ee
Although this result is obtained for $d\geq 4$, an exact computation for $d=2,3$ shows that it actually holds for all $d \geq 2$.
In particular for $d=2$, (\ref{localleading}) gives $S_A=\f{p}{3}\log\frac{L}{a}$ (after an infinite constant term is dropped).
Therefore all theories with $f(x)=x^p$ are local and obey the area law.

Now we make the theory nonlocal by choosing $f(x)=e^{x^q}$. For all $S^d$ with $d\geq 2$, the leading divergence becomes
\be
\label{field}
S_A =\f{2q}{6\cdot
(d-2)! \cdot (d-2+2q)}~\left(\frac{L}{a}\right)^{d-2+2q}.
\ee
Therefore $S_A$ obeys the volume law when $q=\f{1}{2}$, in any dimension $d$.
To summarize, we find that a nonlocal scalar field theory defined by the
action
\be S_{boundary}=\int d\Omega_d \left[\phi \cdot
e^{\s{-\Delta}}\cdot \phi\right], \label{actionss} \ee has entanglement entropies that exhibit the volume law.

The bulk Bekenstein bound requires that the maximal boundary \emph{statistical} entropy is bounded by the volume law. We have computed, for the ground state in the Euclidean theory on $S^d$, the entanglement entropy when \emph{exactly} half of the system is traced out. This directly translates into the statistical entropy of the Minkowski theory on $dS_d$, thus the boundary statistical entropy for the ground state satisfies the volume law.

For thermal states we simply replace the flat space with Schwarzschild black holes. The entanglement entropy computed holographically still obeys the volume law. Indeed, since the volume law behavior stems from the nonlocal nature of the boundary theory, it cannot be changed by simply exciting the system. Hence the maximal boundary statistical entropy is bounded by the volume law, which ensures the bulk Bekenstein bound.

We have used a free scalar theory to show that a nonlocal theory is needed to realize the  volume law for entanglement entropies. The full holographic dual of flat space is likely to contain more fields and to be strongly-interacting. As in AdS/CFT and in standard entanglement entropy computations, adding more fields and turning on local interactions do not alter the scaling behavior of the entanglement entropy; we expect that appropriate nonlocal interactions preserve the volume law as well.

We speculate that the holographic dual of the flat
space is given by the nonlocal generalization of a non-abelian
gauge theory on $S^d$: the theory now has a nonlocal kinetic term
like (\ref{actionss}). Indeed, a similar nonlocal structure is
known to appear in open string field theory (see e.g.\cite{SFT}).
Moreover, the unconventional kinetic term in (\ref{actionss}) is
natural when we rewrite our flat space metric into
$ds^2=\f{dr^2}{r^2}+(\log r)^2 d\Omega^2_d$ and
draw a parallel with AdS metric
$ds^2=\f{dr^2}{r^2}+r^2d\vec{x}^2$: the boundary kinetic
terms of these two spaces scale the same since $e^{\de_{\Omega}}
\sim r  \sim \de_x $. This comparison also shows that
$\rho=\log r$ should be regarded as the energy scale, hence
$\rho_\infty$ is the UV energy cutoff --- corresponding to the UV cutoff
from the viewpoint of open string theory \cite{PP}. This argument
can be seen as a logarithmic generalization of the holographic
correspondence of Lifshitz-like fixed points introduced in
\cite{Lif}.

\noindent
{\bf Holographic correlation functions.}
Another important quantity in establishing the holography is the
correlation function. Now we extend the bulk-to-boundary procedure \cite{GKP} to the flat space (\ref{metric}) and compute its holographic correlation functions.

Consider a scalar field theory in the bulk:
\be S_{bulk}=\f{1}{32\pi G^{(d+1)}_N}\int
d^{d+1}x
\s{g}~(\de_\mu\phi\de^\mu\phi+M^2\phi^2). \label{massless}\ee
Under the Dirichlet boundary condition, the boundary/bulk correspondence is $
\langle e^{\int d\Omega_d \Phi\cdot \hat{O}}\rangle_{S^d}=Z_{\mathbb{R}^{d+1}}[\phi(\rho_{\infty},\Omega)=\Phi(\Omega)]$,
where the left-hand side is the generating functional of correlation functions on the boundary $S^d$ and the right-hand side is the bulk partition function under the Dirichlet boundary condition; $\Phi$ is the boundary scalar field, which sources the operator $\hat{O}$ in $S^d$. In the classical limit, $Z_{\mathbb{R}^{d+1}}[\Phi]\sim \exp{(-S^{on-shell}_{bulk}[\Phi])}$. 
Since the boundary Newton's constant is $G^{(d)}_N=\frac{G^{(d+1)}_N}{(\rho_{\infty})^{d-1}}$, in the limit of $\rho_{\infty}\rightarrow \infty$ the boundary theory decouples from gravity.

For the massless scalar, $\phi(\rho,\Omega)$ can be solved using the bulk Dirichlet Green's function. The on-shell action gives the boundary two-point function:
\be
\la
\hat{O}(\Omega_1)\hat{O}(\Omega_2)
\lb=\frac{1}{2^{\frac{d+9}{2}}\pi\mathcal{A}_{d}}\cdot\f{\rho_{\infty}^{d-1}}{G_N^{(d+1)}}\f{1}{(1-\cos\theta)^{\f{d+1}{2}}}\label{tmassless}
\ee
where $\cos\theta=\vec{\Omega}_1\cdot\vec{\Omega}_2$ and $\mathcal{A}_{d}$ is the area of the unit sphere $S^d$. This agrees with the analysis in
\cite{Solo}, though our interpretations are slightly different.
Since (\ref{tmassless}) contains only a divergent
term, the physical two-point function vanishes after nonlocal boundary counterterms are added to cancel the divergence. Nonlocal counterterms were also used in the holography of
NS5-branes \cite{MiSe,Marolf}.

For the massive scalar, we decompose the boundary field as: $\Phi(\Omega)=\sum_{l,\vec{m}}c_{l,\vec{m}}Y_{l,\vec{m}}(\Omega)$, where $Y_{l,\vec{m}}(\Omega)$ are the orthonormal
spherical harmonics on $S^d$. 
Then the same bulk-to-boundary procedure produces the boundary
two-point function:
\be
\la\hat{O}(\Omega_1)\hat{O}(\Omega_2)\lb
=\frac{\rho^d_\infty}{32\pi G^{(d+1)}_N}\sum_{l,\vec{m}}F_l(\rho_\infty)
\bar{Y}_{l,\vec{m}}(\Omega_1)Y_{l,\vec{m}}(\Omega_2)
\label{tmassive} \ee where $F_l(\rho)\equiv\partial_{\rho}\log{\left(\rho^{\f{1-d}{2}}I_{l+\f{d-1}{2}}(M\rho)\right)}$ and $I_\nu(z)$ is the modified Bessel function of the first kind.

$\rho^d_{\infty} F_l(\rho_{\infty})$ as a polynomial of $\rho_{\infty}$ has a nonzero constant term, which is a degree $[\frac{d}{2}]$ polynomial of $l(l+d-1)$ (eigenvalue of $-\Delta$).
Using the identity
$\sum_{l,\vec{m}}\bar{Y}_{l\vec{m}}(\Omega_1)Y_{l\vec{m}}(\Omega_2)=\delta(\Omega_1-\Omega_2)$, we see that in the limit $\rho_\infty\to\infty$, after counterterms are added to cancel the divergence, the two-point functions consist of $\delta$-functions and their derivatives by Laplacian.  Therefore, the holographic correlation functions for a massive scalar are essentially zero.

Next one could explicitly compute higher-point functions following the bulk-to-boundary principle. However, if we assume the dilaton-type massless scalar Lagrangian of the form ${\cal
L}=(\de_\mu\phi)^2P(\phi)$ where $P(\phi)$ is a polynomial, the correlators always scale as $\rho^{d-1}_\infty$. Therefore they can all be eliminated by adding boundary counterterms.

In summary, we argue that all $n$-point correlation functions vanish after counterterms are added to cancel the
divergences. This seems surprising until one recalls our previous observation from the holographic entanglement entropy: $A$ is maximally entangled with $B$ when the size of $A$ approaches zero.
Define an infinitesimal subsystem $A$ as the disjoint union of the
$n$ points in the correlation function: $A=\sqcup^{n}_{i=1} x_{i}$. Our previous result implies that in this case the entanglement entropy $S_A$ is maximal, therefore the density matrix $\rho_A$
factorizes into a direct product
$\rho_A=\otimes^{n}_{i=1}\rho_{x_{i}}$ where $\rho_{x_i}$ gives the maximal entropy for the subsystem at point $i$, as in a system at an infinitely high temperature. Therefore all correlation functions vanish:
\be \la \hat{O}(x_1)\ldots \hat{O}(x_n)\lb \equiv \mbox{Tr}[\rho_A
\hat{O}(x_1)\ldots \hat{O}(x_n)]=0. \ee
We emphasis that this does not mean that the boundary theory is empty: it stems from the fact that the boundary theory is nonlocal and highly entangled. Based on this we propose that the bulk physics in flat space should be reproduced by the boundary entanglement entropies (with all possible subsystems traced out).


The correlators of the free scalar toy model (\ref{actionss}) are not exactly zero, but the usual divergence (when two operators coincide) already disappears. This leads us to expect that choosing appropriate interactions can further reduce the correlators to zero. Indeed, such a theory already exists in the discretized form: consider a spin model with the randomized antiferromagnetic Heisenberg interaction $H=J\sum_{\langle i,j\rangle}\vec{\sigma}_i\cdot \vec{\sigma}_{j}$ where $J>0$ and $\langle i,j\rangle$ are pairs of two randomly chosen spins. Since the distance and orientation between two spins inside a pair is randomly distributed, all correlators are zero; and since generically the two spins in a given pair are separated far away from each other, the entanglement entropy obeys the volume law. The continuum limit of this type of spin models would provide field theory candidates for the holographic dual of the flat space.

\noindent
{\bf Discussion}.
Now we draw an analogy between the boundary entanglement entropy and the bulk Schwarzschild entropy in the spirit of the connection between Unruh effect and Hawking radiation.
In the Lorentzian version of (\ref{metric}) given by
$ds^2=d\rho^2+\rho^2(-dt^2+\cosh^2 t ~ d\Omega^2_{d-1})$, a static observer at $\rho=\rho_0$
detects a thermal state at the Unruh temperature
$T_U=\f{1}{2\pi\rho_0}$. The entanglement entropy $S_A$ for maximal size $A$
($\ap=\pi$) can be rewritten as $S_A=\f{2^{-d-1}\pi^{\frac{1-d}{2}}}{\Gamma(\frac{d+1}{2})\cdot
G^{(d+1)}_N \cdot T_U^{d-1}}$.
Since $S_A$ measures the amount of
information hidden in the subsystem $B$, which is inaccessible to observers in $A$,
it is analogous to the entropy of Schwarzschild
black hole with temperature $T_{BH}=T_U$. Indeed, the $(d+1)$-dimensional Schwarzschild
black hole has an entropy $S_{BH}=\frac{2^{-2d+1}\pi^{\f{2-d}{2}}(d-2)^{d-1}}{
\Gamma(\frac{d}{2})\cdot{G^{(d+1)}_N \cdot T_{BH}^{d-1}}}$,
which agrees with $S_A$ up to a numerical constant. Thus our holographic
interpretation is consistent with black hole entropies.

This also suggests a string theory interpretation of our
holography. In AdS/CFT \cite{Maldacena}, the holographic dual
theory comes from the D-branes that originally sit at the
horizon $r=0$. In our flat spacetime, in the limit of $\rho\to 0$,
the Unruh temperature $T_U$ becomes infinitely large and the
corresponding observer detects pair creations of many D-branes.
We speculate that their open string theory is
the nonlocal field theory conjectured to be the
holographic dual of the flat spacetime.

Finally, let us reexamine the connection between UV cutoff in the field theory and the cutoff radius of the bulk. Matching the entanglement entropy obtained from the holographic computation ((\ref{entt}) with $\alpha=\pi$) and the field theory one ((\ref{field}) with $q=\frac{1}{2}$), we see that if we switch to dimensionless coordinates defined by $\tilde{\rho}\equiv \frac{\rho}{R}$ where $R$ is a length unit, and accordingly consider the boundary theory on a unit sphere $S^d$ with dimensionless lattice spacing $\tilde{a}=\frac{a}{R}$, we can identify $\tilde{a}=\frac{1}{\tilde{\rho}_{\infty}}$.
Now $S_A$, interpreted as the entanglement entropy
for the holographic dual on the unit sphere $S^d$, scales as:
$S_A
\sim \frac{n}{\tilde{a}^{d-1}}$, where the dimensionless number $n\sim \frac{R^{d-1}}{G^{(d+1)}_N}$ counts the number of
fields in the holographic dual. Since the bulk metric $ds^2=R^2(d\ti{\rho}^2+\ti{\rho}^2d\Omega^2_d)$ is
invariant under the rescaling $(R,\ti{\rho})\rightarrow (R\lambda,\ti{\rho}/\lambda)$ for arbitrary $\lambda$, there exists a corresponding symmetry in the holographic dual theory:
$(n,\tilde{a})\to
(\lambda^{d-1}n,\lambda \tilde{a}).$
Note that the total number of degrees of freedom in the boundary field theory is proportional to $
\frac{n}{\tilde{a}^{d-1}}$ and therefore remains invariant. This symmetry suggests
that the theory is highly nonlocal and entangled and will be useful
when we go on to identify the precise holographic dual. We leave
these questions for future study.

\noindent
{\bf Acknowledgement}.
Authors are supported by WPI Initiative, MEXT, Japan.
TT is supported in part by JSPS Grant-in-Aid for Scientific Research No.\ 20740132 and
JSPS Grant-in-Aid 
for Creative Scientific Research No.\ 19GS0219.


\end{document}